\documentclass[aps,jcp,twocolumn,superscriptaddress,floatfix,10pt]{revtex4}
\usepackage{graphicx}
\usepackage{amsmath}
\usepackage{color}
\usepackage{latexsym}
\usepackage{comment}

\begin{document}

\title{Thermal conductivity of commodity polymers under high pressures}

\author{Otavio Higino Moura de Alencar}
\affiliation{CentraleSup\'elec, Universit\'e Paris--Saclay, 91190 Gif--sur--Yvette, France}
\affiliation{Centro de Tecnologia, Universidade Federal do Cear\'a, 60455-760 Fortaleza, Cear\'a, Brazil}
\author{James Wu}
\affiliation{Quantum Matter Institute, University of British Columbia, Vancouver BC V6T 1Z4, Canada}
\affiliation{Department of Physics, University of California, Berkeley, California 94720, United States}
\author{Marcus M\"uller}
\affiliation{Institut f\"ur Theoretische Physik, George--August--Universit\"at G\"ottingen, 37077 G\"ottingen, Germany}
\author{Debashish Mukherji}
\email[]{debashish.mukherji@theorie.physik.uni-goettingen.de}
\affiliation{Quantum Matter Institute, University of British Columbia, Vancouver BC V6T 1Z4, Canada}
\affiliation{Institut f\"ur Theoretische Physik, George--August--Universit\"at G\"ottingen, 37077 G\"ottingen, Germany}


\begin{abstract}
Understanding the thermal conductivity of polymers under high--pressure conditions is essential for a range of applications, 
from aerospace and deep--sea engineering to common lubricants. However, the complex relationship between pressure, $P$, 
the thermal transport coefficient, $\kappa$, and polymer architecture poses substantial challenges to 
both experimental and theoretical investigations.
In this work, we study the pressure--dependent thermal transport properties of a widely used commodity polymer -- 
poly(methyl methacrylate) (PMMA) -- using a combination of all--atom molecular dynamics 
simulations and semi-analytical approaches. While we report both classical and quantum-corrected estimates of $\kappa$, 
the latter approach reveals that as the pressure increases from 1 atm to 10 GPa, $\kappa$ rises by up to a 
factor of four -- from 0.21 W m$^{-1}$ K$^{-1}$ to 0.80 W m$^{-1}$ K$^{-1}$.
To better understand the mechanisms behind this increase, we disentangle the contributions from bonded and 
nonbonded monomer interactions. Our analysis shows that nonbonded energy-transfer rates increase by a factor of six 
over the pressure range, while bonded interactions show a more modest increase -- about a factor of three. This
observation further consolidates the fact that the nonbonded interactions play the dominant role in dictating the 
microscopic heat flow in polymers.
These individual energy-transfer rates are also incorporated into a simplified heat diffusion model to predict $\kappa$.
The results obtained from different approaches show internal consistency and align well with available experimental data.
Additionally, some data for polylactic acid (PLA) are presented.
\end{abstract}

\maketitle

\section{Introduction}
\label{sec:intro}

Polymers are ubiquitous in our every life due to their lightweight nature, mechanical flexibility, chemical tunability, 
and cost-effectiveness~\cite{DGbook}. 
Due to their spatially extended, fractal configurations~\cite{ZGWang17Mac,kremer23as}, their applications span a wide range of 
technologies, including aerospace components~\cite{WRIGHT1991222}, deep-sea materials~\cite{deepSeaRev}, and automotive lubricants~\cite{GAO2020273,kappaOil}. 
In many of these settings, polymers are subjected to demanding environmental conditions, particularly elevated pressures and temperatures, 
which impose stringent requirements on thermal stability and heat dissipation.
However, the intrinsically low thermal conductivity of polymers -- quantified by the thermal transport coefficient $\kappa$ -- 
remains a key limitation in such applications, often restricting their performance in heat-sensitive or high-power environments.

Heat transport in polymers arises from a complex interplay between bonded interactions along the polymer backbone (e.g., carbon-carbon (C-C) bonds), 
nonbonded interactions (e.g., van der Waals forces or hydrogen bonds), and the overall polymer conformation~\cite{PolRevTT14,Keblinski20,DM24rev}.
In neutral amorphous polymers, where local (monomer level) vibrations dominated by nonbonded interactions carry heat, 
$\kappa$ typically ranges from 0.1 to 0.4 Wm$^{-1}$K$^{-1}$~\cite{Cahill16Mac,DM24rev}.
In contrast, when bonded interactions become dominant -- such as in highly aligned or stretched polymer fibers -- 
$\kappa$ can increase significantly, reaching values between 50--100 Wm$^{-1}$K$^{-1}$ for polyethylene fibers~\cite{shen2010polyethylene,cahill13mac}
and 14 Wm$^{-1}$K$^{-1}$ for cellulose fibers~\cite{CNF2022}.
%

While thermal conductivity of polymers has been studied extensively under ambient conditions~\cite{PolRevTT14,Keblinski20,DM24rev}, 
relatively little is known about their behavior at elevated pressures~\cite{HighPPrivalko01111989,HighPPrivalko1992,Andersson1994,HighPDAWSON2006268,CahillHighP}. 
High pressure is expected to exert a pronounced influence on heat flow, as chains can densify, reorganize, or undergo phase transitions -- 
all of which can markedly influence $\kappa$.
The nontrivial coupling between pressure-induced structural changes and the different modes of energy 
transfer in polymers make this a challenging yet important area of study.

In this work, we employ all-atom molecular dynamics simulations to investigate the pressure dependence 
of thermal conductivity in two representative commodity polymers: poly(methyl methacrylate) (PMMA) and polylactic acid (PLA).
The macroscopic $\kappa$ is computed using a standard nonequilibrium approach-to-equilibrium method~\cite{Lampin2013},
and the quantum-corrected values are estimated using an approach that properly accounts for the
vibrational modes contributing to $\kappa$ at a given temperature and pressure~\cite{Mukherji24PRM}. To gain microscopic insight, we utilize 
a single-chain energy-transfer model~\cite{MM21acsn,MM21mac} to examine how pressure modulates 
energy transfer between monomers, both bonded and nonbonded, and how these microscopic changes affect the macroscopic $\kappa$.
By decoupling these contributions and embedding them within a simplified diffusion model, we aim to uncover the underlying mechanisms 
that govern thermal transport in polymers under extreme pressures. This understanding may offer valuable insights for designing 
materials optimized for high-pressure applications.

The remainder of this manuscript is organized as follows:
Section~\ref{sec:ModMeth} introduces the computational method, including 
the relevant material-specific details.
Section~\ref{sec:res} presents the results.
Finally, conclusions are drawn in Section~\ref{sec:conc}.

\section{Materials, models and methods}
\label{sec:ModMeth}

\subsection{Polymer models}

As test cases, we simulate two widely used commodity polymers: poly(methyl methacrylate) (PMMA) and polylactic acid (PLA). 
While both polymers are of interest, our analysis will primarily focus on PMMA due to the extensive availability of experimental data for comparison.
PMMA is modeled using a modified version of the OPLS-AA force field, as described in Ref.~\cite{Mukherji17NC}, 
while PLA is represented using the standard OPLS-AA parameters \cite{OPLS}. These force field choices have been previously validated, 
accurately capturing the conformational behavior of single PMMA chains in solution~\cite{Mukherji17NC}, as well as 
reproducing the heat capacities of PMMA~\cite{MHM21prm} and PLA~\cite{Mukherji24PRM}, and the thermal conductivity of both polymers~\cite{Mukherji24PRM}.

Each system consists of $N_c = 100$ polymer chains, with each chain composed of $N_{\ell} = 30$ monomer units.
In this work, we utilize previously equilibrated, solvent-free PMMA~\cite{MHM21prm} and PLA~\cite{Mukherji24PRM} configurations.
All systems were initially equilibrated in their melt state at a temperature $T = 600$ K under ambient pressure (1 atm), and subsequently quenched down to $T = 300$ K. Further methodological details are presented in the corresponding sections.

\subsection{Simulation details}

All simulations are performed using the GROMACS molecular dynamics package~\cite{Abraham:2015}.
While the initial configurations are taken from our previous studies, in the present work we re-equilibrate the systems at 
each target pressure $P$ at a fixed temperature of $T = 300$ K. Temperature is controlled using 
a Langevin thermostat with a time constant $\tau_{\rm T} = 1.0$ ps.
Pressure $P$ is varied from 1 atm to 10 GPa, and is maintained using the Parrinello-Rahman barostat with a time constant $\tau_{\rm p} = 0.5$ ps.
These simulations are performed for 1 $\mu$s at each $P$, where the integration time is chosen to be $\Delta t = 1$ fs.
After equilibration at each $P$, we compute key observables -- including the thermal conductivity $\kappa$, monomer-level energy-transfer rates, 
and elastic constants $C_{ij}$ -- all at $T = 300$ K.

\section{Results and discussions}
\label{sec:res}


\begin{figure}[ptb]
\includegraphics[width=0.49\textwidth,angle=0]{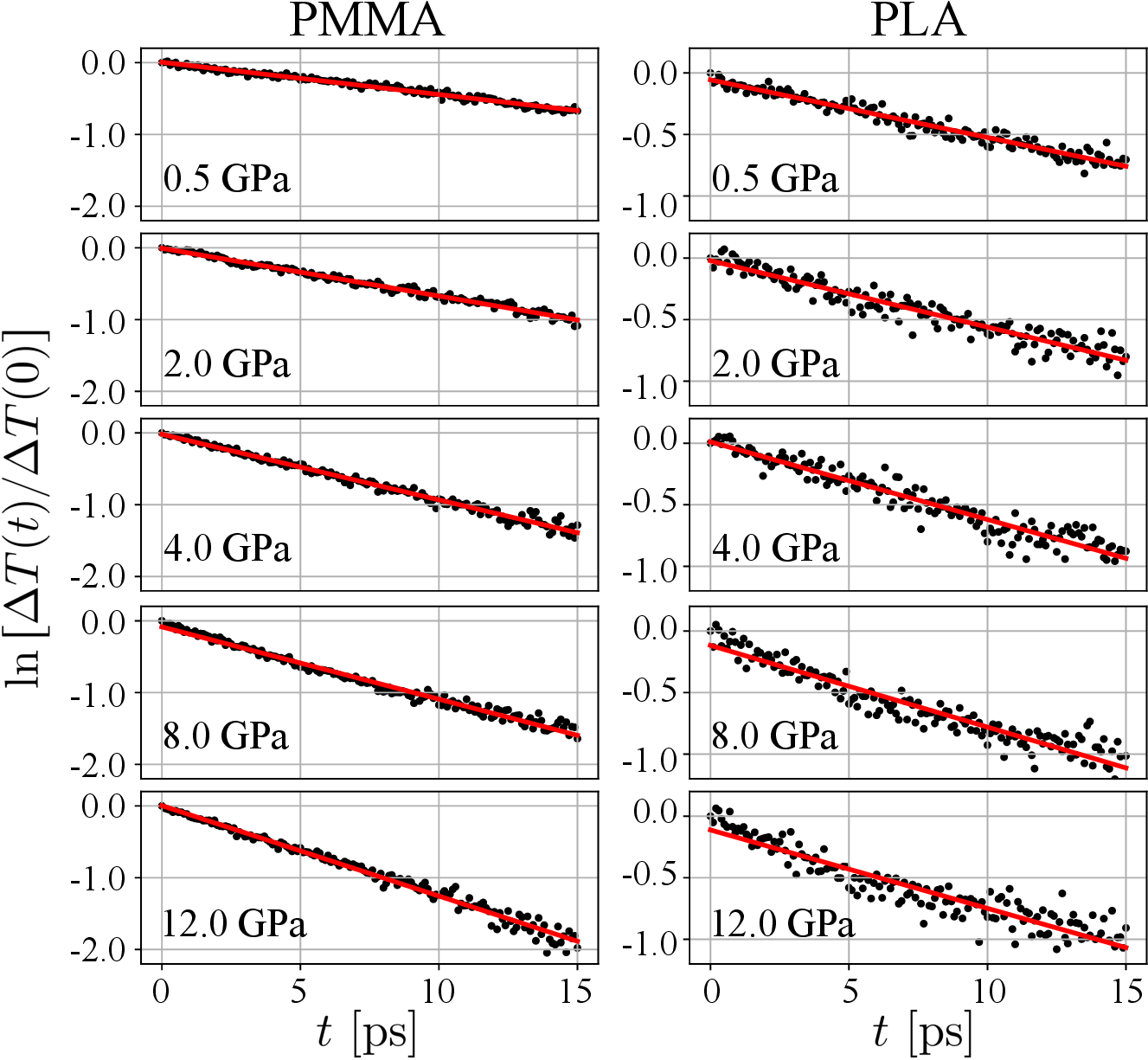}
\caption{Time evolution of the kinetic temperature difference between hot and cold slabs,
$\Delta T(t) = T_{\rm hot}(t) - T_{\rm cold}(t)$, for (left) poly(methyl methacrylate) (PMMA) and (right) polylactic acid (PLA).
Results are shown for different pressures $P$. Solid lines represent exponential fits,
from which the thermal conductivity $\kappa$ is extracted using Equation~\ref{eq:ate}.
Note that slightly larger fluctuations in the PLA data is because it consists of a smaller number of atoms within the simulation domain
owing to its smaller monomer size.}
\label{fig:taupol}
\end{figure}

We begin by computing the classical estimates of the thermal conductivity coefficient $\kappa$ as a function of applied pressure $P$.
For this purpose, we employ the nonequilibrium approach-to-equilibrium method~\cite{Lampin2013}, 
which we have previously validated across a range of polymeric systems~\cite{MM21mac,Mukherji24PRM}.
In this method, the simulation box is divided into two equal regions: one initialized at a higher 
temperature ($T_{\rm hot} = 350$ K), and the other at a lower temperature ($T_{\rm cold} = 250$ K). 
These temperatures are imposed using a Langevin thermostat applied for 5 ns with a timestep of $\Delta t = 1$ fs.
After this initial thermalization, the thermostat is removed, and the temperature difference $\Delta T(t) = T_{\rm hot}(t) - T_{\rm cold}(t)$ 
is allowed to relax in the microcanonical (NVE) ensemble. The NVE simulations are carried out for 15 ps with a reduced timestep of $\Delta t = 0.1$ fs 
to ensure numerical stability. During this relaxation period, $\Delta T(t)$ decays naturally as heat flows from the hotter to the colder region.
Figure~\ref{fig:taupol} shows the transient decay of $\Delta T(t)$ for both polymers across increasing applied 
pressures $P$ (from top to bottom). The decay is well described by an exponential function, $\Delta T(t) \propto \exp(-t/\tau)$, 
where $\tau$ is the characteristic relaxation time extracted from the fits (shown as red lines).

The relaxation times $\tau$ extracted from the temperature decay curves in Figure~\ref{fig:taupol} can be used to compute 
$\kappa$ via the following relation,
\begin{equation}
\kappa = \frac{1}{4\pi^2} \frac{c L}{{\mathcal A} \tau},
\label{eq:ate}
\end{equation}
where $c = 3 N k_{\rm B}$ is the Dulong-Petit classical estimate of specific heat, $L$ is the length of the simulation domain 
along the direction of heat flow, and $\mathcal A$ is the cross-sectional area perpendicular to that direction.
Here, $N$ is the total number of atoms in the systems and $k_{\rm B}$ is the Boltzmann constant.

\begin{figure}[ptb]
\includegraphics[width=0.49\textwidth,angle=0]{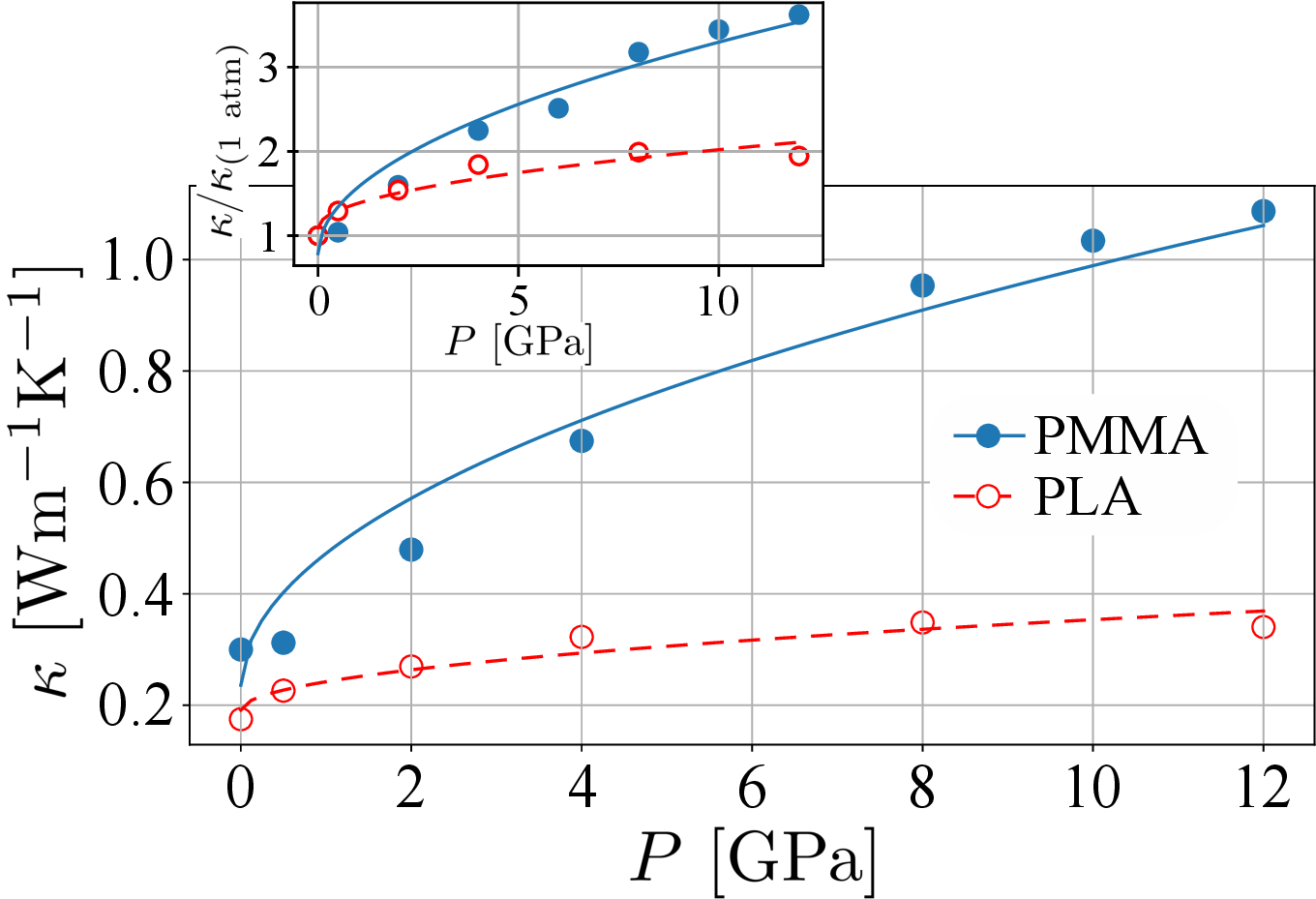}
\caption{The main panel shown the classical estimate of the thermal transport coefficient $\kappa$ as a 
function of pressure $P$ for two polymers: poly(methyl methacrylate) (PMMA) and polylactic acid (PLA), at $T = 300$ K. 
The lines represent fits to the empirical relation $\kappa = \kappa_1 + \kappa_2 \sqrt {P}$.
The inset shows the same data as the main panel but with $\kappa$ normalized by its value at $P = 1$ atm, 
i.e., $\kappa/\kappa_{\mathrm{(1~atm)}}$, for the respective PMMA and PLA simulations.}
\label{fig:kappa_p}
\end{figure}

As expected, the relaxation of $\Delta T(t)$ accelerates progressively with increasing $P$, indicating enhanced heat transfer 
under compression. This trend is reflected in the pressure-dependent increase of $\kappa$, 
as shown in the main panel of Figure~\ref{fig:kappa_p}, and is consistent with previous observations of pressure-enhanced thermal 
transport in polymers~\cite{HighPPrivalko01111989,HighPPrivalko1992,Andersson1994,HighPDAWSON2006268,CahillHighP}.
Quantitatively, as pressure increases from 1 atm to approximately 12 GPa, $\kappa$ increases by a factor of about 2.0 for 
PLA and around 3.5 for PMMA (see the inset in Figure~\ref{fig:kappa_p}). 
The observed increase in $\kappa$ can be interpreted within the framework of the minimum thermal conductivity model~\cite{Cahill90PRB,CahillHighP}, 
which connects $\kappa$ to both the material’s stiffness and atomic number density $\rho_N$. 
Under increasing $P$, the stiffness (i.e., the elastic moduli) tends to grow approximately linearly with $P$, 
while $\rho_N$ remains relatively unchanged. This results in an empirical scaling of the form $\kappa = \kappa_1 + \kappa_2 \sqrt {P}$,
which has been previously validated in experimental studies~\cite{CahillHighP} and is used here to fit our simulation data of classical $\kappa$ using Equation~\ref{eq:ate}.
%
%
Our simulation results are well described by this empirical model, as evidenced by the fitted curves in Figure~\ref{fig:kappa_p}. 
The best fit parameters obtained from the simulations are $\kappa_1 = 0.191$ Wm$^{-1}$K$^{-1}$ and $\kappa_2 = 0.051$ Wm$^{-1}$K$^{-1}$GPa$^{-1/2}$ 
for PLA, and $\kappa_1 = 0.239$ Wm$^{-1}$K$^{-1}$ and $\kappa_2 = 0.234$ Wm$^{-1}$K$^{-1}$GPa$^{-1/2}$ for PMMA. 
At a later stage in this work, we will come back to the effect of $P$ on stiffness of PMMA in a more detail.

We note in passing that the absolute values of $\kappa$ obtained from experiments~\cite{Andersson1994,CahillHighP} are typically
smaller than those estimated from our classical simulations in Figure~\ref{fig:kappa_p}. A key contribution to this discrepancy is the overestimation of the 
specific heat capacity $c$ in classical models. 
Classical simulations assume that all vibrational modes contribute equally to $c$, whereas in reality, many 
high-frequency modes in polymers -- such as C-H bond vibrations with characteristic frequencies around $\nu = 90$ THz --
remain quantum mechanically frozen at room temperature ($T = 300$ K, corresponding to approximately $\nu = 6.2$ THz)~\cite{MHM21prm}.
When $\kappa$ is corrected by incorporating the appropriate contributions from the vibrational density of states at a 
given temperature, the simulation results show good agreement with experimental measurements~\cite{Mukherji24PRM}.

The discussions in the preceding paragraphs invoke a few fundamental questions:\\
(i) What is the role of the increased atomic number density $\rho_N$ with pressure $P$ in governing monomer-level energy-transfer rates?\\
(ii) How does the monomer-level energy transfer influences macroscopic $\kappa$?\\
(iii) How does the interplay between $\rho_N$ and the inherent increase in material stiffness with $P$ influence the thermal conductivity $\kappa$?\\
(iv) Can quantum-corrected $\kappa$ data offer deeper insight into the underlying physical mechanisms?\\
To address these questions, we employ three recently proposed techniques: (1) the single-chain energy-transfer model~\cite{MM21acsn,MM21mac}, 
(2) a noise-canceling method for accurately estimating the components of the elastic modulus tensor $C_{ij}$~\cite{Herzbach06cpc,Muser25sub}, 
and (3) a framework for calculating quantum-corrected $\kappa$ using the exact vibrational density of states $g(\nu)$~\cite{Mukherji24PRM}, 
under the assumptions of the minimum thermal conductivity model~\cite{Cahill90PRB}.\\
In the following sections, we apply these approaches to gain deeper insight into thermal transport under 
high-pressure conditions. Given the availability of detailed experimental data for PMMA, 
our analysis will primarily focus on this system.

\subsection{Effect of pressure on monomer-level energy transfers}

Macroscopic $\kappa$ in polymers is typically low. However, at the microscopic (monomer) level, 
energy transfer occurs through distinct pathways: (i) between two bonded monomers, and (ii) between 
a monomer and its $n$ nonbonded nearest neighbors.
For instance, a typical bonded interaction in polymers is the C-C covalent bond, which has a bond 
strength of approximately $80k_{\rm B}T$ at $T = 300$ K and a stiffness exceeding 250 GPa~\cite{CCBondE}. 
In contrast, nonbonded interactions in neutral polymers are primarily governed by van der Waals and hydrogen bonds, 
with interaction strengths ranging from 1 to 4 $k_{\rm B}T$~\cite{Mukherji20AR,desiraju02}.
As a result of these much stiffer bonded interactions, heat transfer between covalently bonded monomers 
is significantly faster than between nonbonded monomers. A model capable of decoupling these two 
distinct microscopic energy-transfer rates along these pathways is the single-chain energy-transfer model (SCETM)~\cite{MM21acsn,MM21mac}.

SCETM is based on a simplified yet effective picture in which an energy packet primarily diffuses along the polymer backbone via stiff
bonded interactions through successive hops. Occasionally, the energy also transfers off the chain 
to nearest-neighbor nonbonded monomers. Within this framework, the time evolution of the internal 
energy ${\mathcal E}_i = c_{\rm mon} T_i$ of the $i^{\rm th}$ monomer can be expressed as,
\begin{align}
\label{eq:scetm}
    \frac {{\rm d} T_i} {{\rm d} t} 
    &=\frac {G_{\mathrm{b}}}{c_{\rm mon}}(T_{i+1}-2T_i+T_{i-1})\\\nonumber
    &\quad+ \frac{\tilde G_{\mathrm{b}}}{c_{\rm mon}}(T_{i+2}-4T_{i+1}+6T_i-4T_{i-1}+T_{i-2})\\\nonumber
    &\quad+\frac{nG_{\mathrm{nb}}}{c_{\rm mon}}(T_{\mathrm{bulk}}-T_i)\,,
\end{align}
where $G_{\rm b}/c_{\rm mon}$ and ${\tilde G}_{\rm b}/c_{\rm mon}$ denote the energy-transfer rates via bonded and next-nearest 
bonded interactions along the chain, respectively, while $G_{\rm{nb}}/c_{\rm mon}$ accounts for the nonbonded 
energy-transfer rate. Here, $c_{\rm mon}$ is the specific heat of a single monomer, and the bulk temperature is taken to be $T_{\mathrm{bulk}} = 300$ K.
It is important to note that Equation~\ref{eq:scetm} explicitly includes only first and second nearest-neighbor bonded 
interactions along the chain backbone, which are typically the dominant contributors to intrachain energy transport.

Following Ref.~\cite{MM21acsn,MM21mac}, diagonalizing Equation~\ref{eq:scetm} along the polymer chain contour yields a set of exponentially 
relaxing temperatures of eigenmodes, 
\begin{equation}
\label{eq:tp}
	{\hat T}_p\left(t\right) \propto e^{-\alpha_p t},
\end{equation}	
where $\hat{T}_p(t)$ represents the $p^{\rm th}$ mode of the cosine-transformed temperature, defined as
\begin{equation}
\label{eq:cos}
	{\hat T}_p (t) = \sum_{i = 0}^{N_{\ell}-1} \left[T_i (t) -T_{\rm bulk}\right] \cos \left[\frac {p\pi}{N_{\ell}}\left(i+ \frac {1}{2}\right)\right],
\end{equation}
and the corresponding relaxation rate $\alpha_p$ is given by,
\begin{equation}
	\label{eq:alphap}	
	\alpha_p = 4 \frac {G_{\rm b}}{c_{\rm mon}}\sin^2\left(\frac {p\pi}{2N_{\ell}}\right) - 16 \frac {{\tilde G}_{\rm b}}{c_{\rm mon}}\sin^4\left(\frac {p\pi}{2N_{\ell}}\right)
	+ n \frac {G_{\rm nb}}{c_{\rm mon}}.
\end{equation}

To calculate the relaxation rates $\alpha_p$, we have used the exact same protocol as in our earlier works~\cite{MM21mac,James22CMS}.
For this purpose, we performed a separate set of targeted simulations using a localized thermal perturbation protocol. 
Specifically, the $16^{\rm th}$ monomer of a single PMMA chain was initialized at an elevated temperature of $T_{\rm 16th} = 1000$ K, 
while all other monomers in the system were maintained at the reference bulk temperature $T_{\rm bulk} = 300$ K. 
This initial thermal configuration was generated through canonical simulations.
Following this thermalization phase, the elevated temperature $T_{\rm 16th}$ was allowed to relax naturally 
during microcanonical simulations, thereby enabling energy transfer from the hot monomer to its bonded neighbors 
along the chain and to nonbonded neighbors from surrounding chains. To ensure statistically reliable results, 
this procedure was repeated for 600 independent realizations by randomly selecting different PMMA chains from the 
homogeneous bulk and initializing velocities using different random seeds. 
The canonical thermalization are performed for 1 ns with $\Delta t = 1$ fs, followed by 20 ps of microcanonical relaxation with $\Delta t = 0.1$ fs. 
The short NVE simulations were sufficient to capture the early-time dynamics of energy redistribution from the initially hot monomer. 

\begin{figure}[ptb]
\includegraphics[width=0.49\textwidth,angle=0]{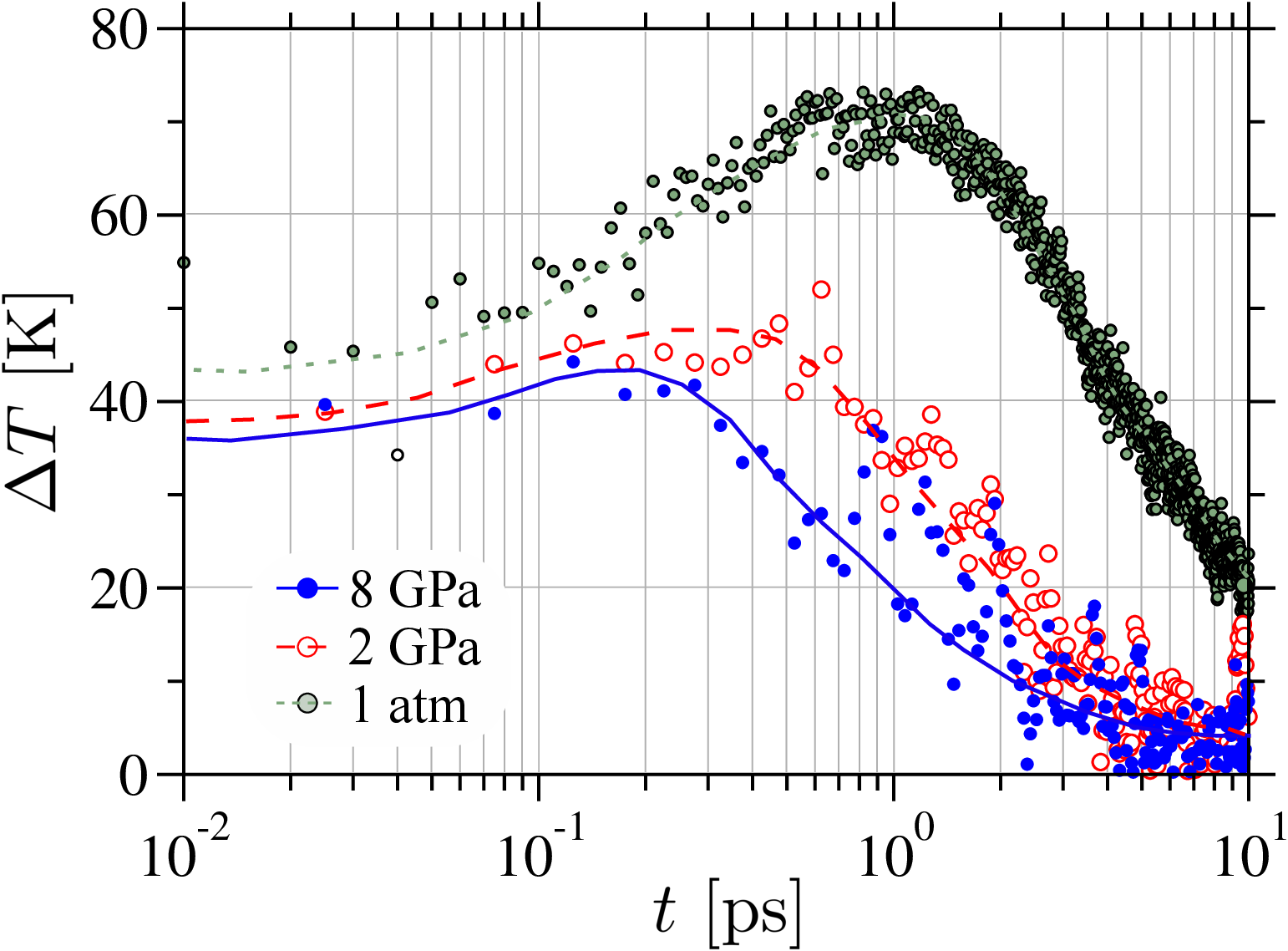}
\caption{Temperature profiles $\Delta T$ as a function of time during the relaxation of a central monomer 
initially heated to 1000 K. The data correspond to its first nearest bonded neighbors, shown for three different pressures $P$. 
Lines are included as visual guides. The data for 1 atm is taken from Ref.~\cite{MM21mac}.}
\label{fig:tempHot}
\end{figure}

In Figure~\ref{fig:tempHot}, we show the temperature evolution of the first bonded neighbor of a hot monomer during NVE simulations. 
As expected, the temperature of this neighbor initially rises as it receives energy from the hot monomer, and subsequently 
decays back to the reference temperature of $T = 300$ K (i.e., $\Delta T = 0$ K). This relaxation occurs as the heat is 
redistributed along the bonded backbone and toward surrounding nonbonded neighbors.
Two features are visible with increasing $P$: (i) the peak shifts towards shorter time and (ii) peak height decreases.
This former trend reflects a faster and more efficient redistribution of energy between monomers under 
compression and can be directly linked to the relaxation rates $\alpha_p$ of the underlying temperature modes. 
Moreover, it also helps explain the latter observation (ii), which is addressed in the following analysis. 

\begin{figure}[ptb]
\includegraphics[width=0.49\textwidth,angle=0]{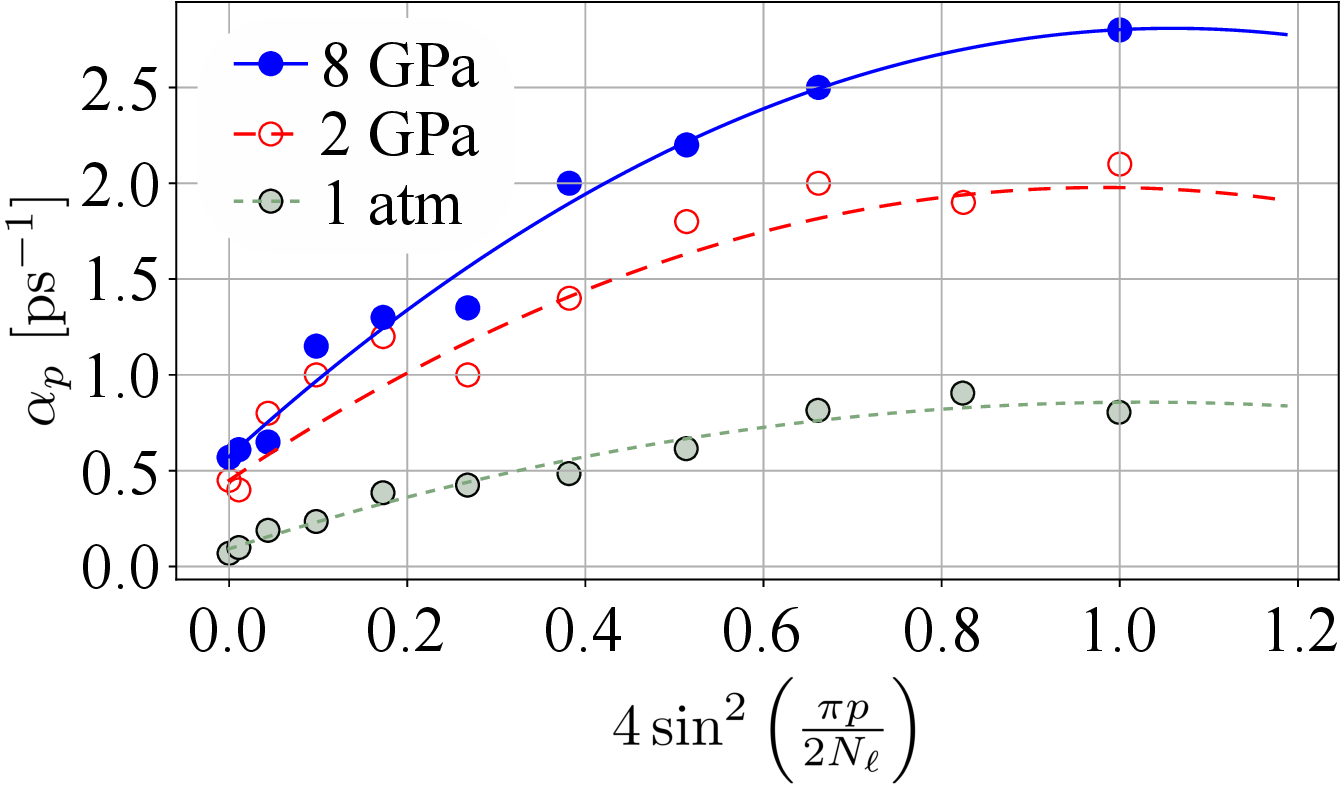}
\caption{The relaxation rates $\alpha_p$ of the cosine-transformed temperature modes $\hat{T}_p(t)$ are plotted 
as a function of $4\sin^2\left(\pi p / 2N_{\ell}\right)$. The data correspond to poly(methyl methacrylate) (PMMA) 
at a temperature of $T = 300$ K, with representative results shown for three different applied pressures $P$. 
The dataset for ambient pressure (1 atm) is taken from Ref.~\cite{MM21mac}.
The solid lines represent fits to the data using Equation~\ref{eq:alphap}.}
\label{fig:alpha_p}
\end{figure}

\begin{table*}[ptb]
        \caption{A table summarizes the energy-transfer rates between different types of monomer pairs, specifically: 
                 nearest neighbor nonbonded monomers $G_{\rm nb}/c_{\rm mon}$, bonded monomers $G_{\rm b}/c_{\rm mon}$, and next-nearest bonded monomers 
                 $\tilde{G}_{\rm b}/c_{\rm mon}$. Additionally, we list the thermal transport coefficients 
                 obtained directly from simulations $\kappa$ alongside the theoretical predictions 
                 based on the SCETM model $\kappa_{\rm theory}$ using Equation~\ref{eq:kappa_heu}.
                 The data are presented for poly(methyl methacrylate) (PMMA) for three different pressure $P$ at the reference temperature $T = 300$ K. }
\begin{center}
       \begin{tabular}{|c|c|c|c|c|c|c|c|c|c|c|c|}
\hline
	       ~~~~$P$~~~~      &   $ {G_{\rm nb}}/{c_{\rm mon}}$ [ps$^{-1}$]        &    $ {G_{\rm b}}/{c_{\rm mon}}$ [ps$^{-1}$]   
                           &   ${{\tilde G}_{\rm b}}/{c_{\rm mon}}$ [ps$^{-1}$] &    $~~~ {G_{\rm b}}/{G_{\rm nb}}~~~$ 
                           &   $\kappa$ [Wm$^{-1}$K$^{-1}$]                     &    $\kappa_{\rm theory}$ [Wm$^{-1}$K$^{-1}$]\\\hline
\hline
	       1 atm            &  0.024     & 1.49   &   0.72 &  61.07   & 0.31 & 0.16 \\   
	       2 GPa            &  0.119     & 3.12   &   1.59 &  26.10   & 0.48 & 0.44 \\   
	       8 GPa            &  0.152     & 4.22   &   1.99 &  27.72   & 0.95 & 0.62 \\   
\hline	       
\end{tabular}  \label{tab:cetm}
\end{center}
\end{table*}

In Figure~\ref{fig:alpha_p}, we present the relaxation rates $\alpha_p$ as a function of the mode index $p$, 
which are well captured by the theoretical prediction provided in Equation~\ref{eq:alphap}. 
The corresponding energy-transfer rates -- specifically for bonded $G_{\rm b}$, next-nearest bonded $\tilde{G}_{\rm b}$, and nonbonded $G_{\rm nb}$ -- 
are extracted from fits to the data and summarized in Table~\ref{tab:cetm}.
A clear and systematic pressure dependence is observed, particularly in the nonbonded energy-transfer channel, 
which are dominated by relatively soft interactions. As pressure is increased from 1 atm to 8 GPa, the nonbonded energy 
transfer rate $G_{\rm nb}$ increases significantly, by approximately a factor of 5–6. In contrast, the energy-transfer 
rate $G_{\rm b}$ between bonded monomers increases more modestly, by a factor of roughly 2–3 
over the same pressure range.
This disparity reflects the different sensitivities of bonded and nonbonded interactions to external pressure. 
Compression leads to a pronounced increase in local packing density, which enhances the frequency and strength of 
intermolecular (nonbonded) contacts. As a result, $G_{\rm nb}$ becomes substantially larger at higher 
pressures, which leads a substantial amount of energy leakage via the nonbonded contacts and 
is the main reason why the peak intensities of $\Delta T$ in Figure~\ref{fig:tempHot} decrease with $P$. 
On the other hand, bonded interactions -- determined by the intrinsic stiffness of covalent bonds -- 
are relatively less effected by compression of up to 10 GPa, leading to a smaller variation in $G_{\rm b}$. 

When we refer to bonded energy transfer, we are referring to the transfers involving all
atoms within a monomer to its neighboring bonded monomer. The effective bond lengths between the center-of-masses 
of two successive monomer are influenced by compression and local packing. In contrast, the lengths of bare 
covalent bonds -- such as C-C bonds (i.e., around 0.15 nm) -- remain essentially unchanged even at pressures up to 10 GPa.

Collectively, the increase in microscopic energy-transfer rates under elevated pressure leads to the observed enhancement 
in the macroscopic thermal conductivity $\kappa$. To quantify this behavior, we utilize a simplified theoretical 
estimate of $\kappa$ previously proposed in Ref.~\cite{MM21mac}, which incorporates contributions from distinct energy-transfer pathways,
\begin{equation}
\label{eq:kappa_heu}
    \kappa_{\rm theory} =\frac{\rho}{6}\left[n G_{\mathrm{nb}}r_{\mathrm{nb}}^2+\left(G_{\mathrm{b}}
    - 4\tilde G_{\mathrm{b}}\right) r_{\mathrm{b}}^2 +\tilde G_{\mathrm{b}} \tilde r_{\mathrm{b}}^2\right],
\end{equation}
Here, $\rho$ is the monomer number density, and $r_{\mathrm{nb}}$, $r_{\mathrm{b}}$, and $\tilde{r}_{\mathrm{b}}$ represent the characteristic 
interaction distances for nonbonded, bonded, and next-nearest bonded monomers, respectively. This expression effectively 
combines the contributions from each energy-transfer mechanism into a single estimate for thermal conductivity.
The calculated values of $\kappa_{\rm theory}$ are summarized in the last column of Table~\ref{tab:cetm}. 
The pressure-dependent increase in $\kappa$ -- typically by a factor of 3 to 4 -- is consistent across all methods used, 
underscoring the robustness of the SCETM-based approach and the simplified diffusion-like model in capturing the essential 
physics of thermal transport under compression.
It is worth noting, however, that the absolute values of $\kappa_{\rm theory}$ tend to underestimate those obtained 
from classical nonequilibrium simulations, see the second last column Table~\ref{tab:cetm}. 
This discrepancy has also been reported in previous studies~\cite{MM21mac,James22CMS} 
across a range of polymers and likely reflects the idealizations inherent in the simplified theoretical framework as well as quantum corrections discussed below.

We are currently unable to pinpoint the precise origin of the discrepancy between $\kappa_{\rm theory}$ and simulation-based 
estimates, which typically ranges between a factor of 1/4 to 2/3 across the various systems studied here as well as in Refs.~\cite{MM21mac,James22CMS}. 
Nor can we fully explain how this deviation is influenced by factors such as polymer stiffness, monomer chemistry, or monomer size.
However, it is worth noting that the simplistic model used to estimate $\kappa_{\rm theory}$ is based primarily on nearest-neighbor 
nonbonded energy transfer, while neglecting cascading or multi-step heat transfer events across extended nonbonded networks. 
This simplification may be significant, as thermal transport in amorphous polymers is often dominated by complex, 
collective interactions involving many-body nonbonded contacts.
A more comprehensive treatment that accounts for these longer-range or higher-order energy-transfer pathways is likely 
needed to improve the quantitative accuracy of $\kappa_{\rm theory}$. A detailed investigation of these mechanisms lies beyond the scope 
of the present work and will be addressed in future studies.

We note in passing that the analysis in Figure~\ref{fig:alpha_p} includes modes up to $p = 11$, 
which corresponds to a spatial resolution of roughly three monomers -- approximately equal to the persistence length of a PMMA chain. 
Below this length scale, energy transport is expected to be ballistic rather than diffusive. 
Consequently, the assumptions underlying the SCETM, which is based on diffusive transport, may no longer be valid in this regime.

\subsection{Quantum estimate of $\kappa$}

So far, we have discussed the microscopic energy-transfer rates and the influence of pressure $P$ on monomer-level 
thermal properties within a classical framework. However, it is important to recognize that in experimental systems 
of amorphous polymers, the macroscopic $\kappa$ is largely governed by low-frequency, vibrational modes -- primarily associated 
with soft, nonbonded interactions. In contrast, the high-frequency modes corresponding to stiff covalent bonds 
(such as C-H or C-C stretching modes) remain quantum-mechanically inactive -- or frozen -- at ambient 
conditions due to their large energy quanta relative to $k_{\rm B}T$~\cite{MHM21prm}.
Classical simulations, however, treat all vibrational modes as fully accessible at all temperatures. 
This results in a systematic overestimation of $\kappa$ in comparison to experimental measurements~\cite{kappaOil,Mukherji24PRM}, 
even when highly accurate force fields are employed and simulations are conducted with meticulous care. 
This fundamental limitation of classical models motivates the need for quantum corrections.

To address this discrepancy, a quantum-corrected approach has been developed based on the original minimum 
thermal conductivity model~\cite{Cahill90PRB}, augmented with the exact vibrational density of states $g(\nu)$, 
obtained from simulation~\cite{Mukherji24PRM}. This method has proven effective in describing thermal 
transport in a wide range of amorphous polymer systems~\cite{Mukherji24PRM}, including cellulose derivatives~\cite{MukherjiSub}.
Within this framework, the thermal conductivity $\kappa(T,P)$ is expressed as
\begin{equation}
    \kappa (T,P) = \left (\frac {\rho_{\rm N} h^2} {6k_{\rm B} T^2}\right)
    \left (v_{\ell}^2 + 2v_{t}^2 \right) 
    \int \frac {\nu e^{{h\nu}/{k_{\rm B}T}}} {\left(e^{{h\nu}/{k_{\rm B}T}} -1 \right)^2}
    g(\nu) {\rm d}\nu,
    \label{eq:mtm}
\end{equation}
where $\rho_{\rm N}$ is the atomic number density, and $v_{\ell}$ and $v_t$ are the longitudinal and transverse sound velocities, respectively. 
These velocities are determined from the elastic moduli and mass density $\rho_{\rm m}$ 
via $v_{\ell} = \sqrt{C_{11}/\rho_{\rm m}}$ and $v_t = \sqrt{C_{44}/\rho_{\rm m}}$.
Equation~\ref{eq:mtm} requires accurate estimates of $\rho_{N}$, $g(\nu)$, and the components of $v_{\rm i}$.
We will now individually compute these quantities and use them to obtain an estimate of the quantum-corrected $\kappa$.

\noindent{\it Vibrational density of states}:
To compute $g(\nu)$ at different pressures, we use the Fourier transform of the normalized mass-weighted velocity 
autocorrelation function $\psi(t)$ using, 
$g(\nu) = \frac {1}{A}\int_{0}^{\infty} \cos(2\pi \nu t) \frac{\psi(t)}{\psi(0)} {\rm d}t$~\cite{Horbach1999JPCB},
where the normalization constant $A$ ensures $\int_0^\infty g(\nu), {\rm d}\nu = 1$. The function $\psi(t)$ is given by,
$\psi(t) = \sum_{i} m_i \langle {\vec{v}}_i(t) \cdot {\vec{v}}_i(0)\rangle$,
and represents the superposition of contributions from all vibrational modes in the system.
For the computation of $\psi(t)$, we use a timestep of $\Delta t = 0.1$ fs and sample data over a 10 ps interval, 
with an output frequency of $5 \times 10^{-4}$ ps. This ensures accurate resolution of both low- and high-frequency vibrational contributions.

\begin{figure}[ptb]
	\includegraphics[width=0.49\textwidth,angle=0]{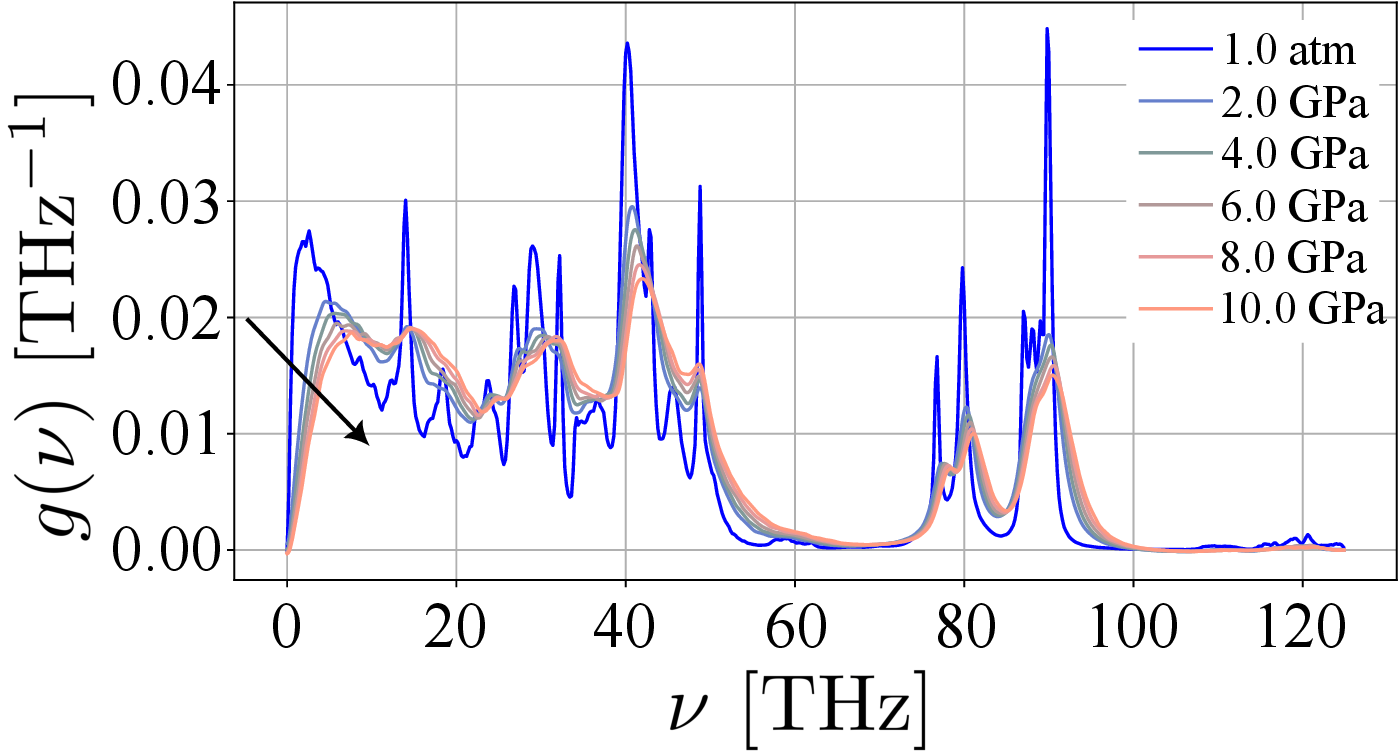}
        \caption{The vibrational density of states $g(\nu)$ for poly(methyl methacrylate) (PMMA) 
        at various pressures $P$, measured at a temperature of $T = 300$ K. 
        The arrow indicates the direction of increasing $P$.
\label{fig:gnu}}
\end{figure}

In Figure~\ref{fig:gnu}, we present representative $g(\nu)$ for PMMA at different $P$.
A prominent feature emerging from $g(\nu)$ is the progressive smoothing of the spectral profile with increasing $P$. 
At ambient conditions, $g(\nu)$ typically exhibits pronounced peaks and irregularities that are a direct consequence 
of the amorphous, disordered nature of polymeric systems. These spectral features arise due to a wide distribution 
of local environments and bonding configurations, particularly in the nonbonded interactions that dominate 
the low-frequency regime. However, as pressure is increased, the local packing density of monomers 
becomes more homogeneous, and the system undergoes a form of pressure-induced structural ordering at the microscopic scale.
This densification under compression reduces the magnitude of thermal fluctuations and local configurational heterogeneity, 
thereby suppressing the vibrational spectrum. Consequently, the sharp peaks and valleys in $g(\nu)$ gradually flatten 
out, and the distribution of vibrational modes becomes more uniform. Additionally, the peaks in $g(\nu)$ shift 
towards higher frequencies with increasing $P$, which is indicative of an overall stiffening of the vibrational modes. 
This is consistent with the enhanced stiffness of PMMA under pressure~\cite{CahillHighP}, particularly in the nonbonded interactions 
where increased packing leads to steeper effective potentials between neighboring monomers.

The smoothing of $g(\nu)$ has further implications for thermal transport. A more uniform vibrational spectrum with 
stiffer modes contributes to a higher thermal conductivity, as energy can be transferred more efficiently through 
vibrational excitations that are less scattered by structural disorder. This observation supports the 
broader conclusion that compression enhances thermal transport in polymers not only through mechanical stiffening, 
but also by reducing vibrational scattering mechanisms inherent to amorphous systems.

\noindent{\it Sound-wave velocities}:
To determine the sound-wave velocities $v_{\ell}$ and $v_t$, the elastic constants $C_{11}$ and $C_{44}$ are first computed 
at finite temperature using a recently proposed method~\cite{Muser25sub}. This method is specifically designed to overcome 
the limitations posed by thermal noise in molecular dynamics simulations. This approach builds on a noise-cancellation 
strategy originally introduced for the calculation of piezoelectric coefficients in crystalline silica~\cite{Herzbach06cpc}, 
and has since been adapted to provide precise estimates of elastic moduli in amorphous commodity polymers~\cite{Muser25sub}
and cellulose derivatives~\cite{MukherjiSub}. 
The key idea is to apply symmetric strain fields and extract stress differences, thereby averaging out stochastic 
thermal fluctuations that typically obscure the small stress responses associated with elastic deformations.

The shear modulus $C_{44}$ is computed by imposing a volume-conserving, anisotropic deformation to an 
initially cubic simulation cell of side length $L$. The deformation is applied such that,
$L_x = L (1 + \varepsilon)$, $L_y = {L}/{(1 + \varepsilon)}$, and $L_z = L$,
ensuring the total volume remains constant to isolate the shear response. 
A small strain $\varepsilon = 10^{-3}$ is used to remain within the linear elastic regime. 
Under these conditions, $C_{44}$ can be computed from the change in stress components using either of the following equivalent expressions;
\begin{equation}
C_{44} = \frac{\sigma_{xx}(+\varepsilon) - \sigma_{xx}(0)}{2\varepsilon}
\quad \text{or} \quad
C_{44} = -\frac{\sigma_{yy}(-\varepsilon) - \sigma_{yy}(0)}{2\varepsilon},
\label{eq:c44}
\end{equation}
where $\sigma_{ij}$ refers to the stress tensor component measured in the deformed configurations. 

The longitudinal modulus $C_{11}$ is extracted by applying uniaxial strains in the $x$-direction in two separate simulations using,
    $L_x = L(1 \pm \varepsilon)$,
while keeping the box lengths along $y$ and $z$ fixed. The corresponding difference in the axial stress response yields;
\begin{equation}
C_{11} = \frac{\sigma_{xx}(+\varepsilon) - \sigma_{xx}(-\varepsilon)}{2\varepsilon}.
\label{eq:c1112}
\end{equation}
This symmetric strain protocol ensures that any nonlinear or temperature-induced offsets are eliminated, 
enabling accurate determination of elastic constants even in thermally active systems.
Overall, this methodology offers a robust 
means of evaluating elastic properties from finite-temperature simulations, 
particularly useful for disordered systems like polymers where standard stress-strain methods often suffer from poor signal-to-noise ratios.

\begin{figure}[ptb]
	\includegraphics[width=0.49\textwidth,angle=0]{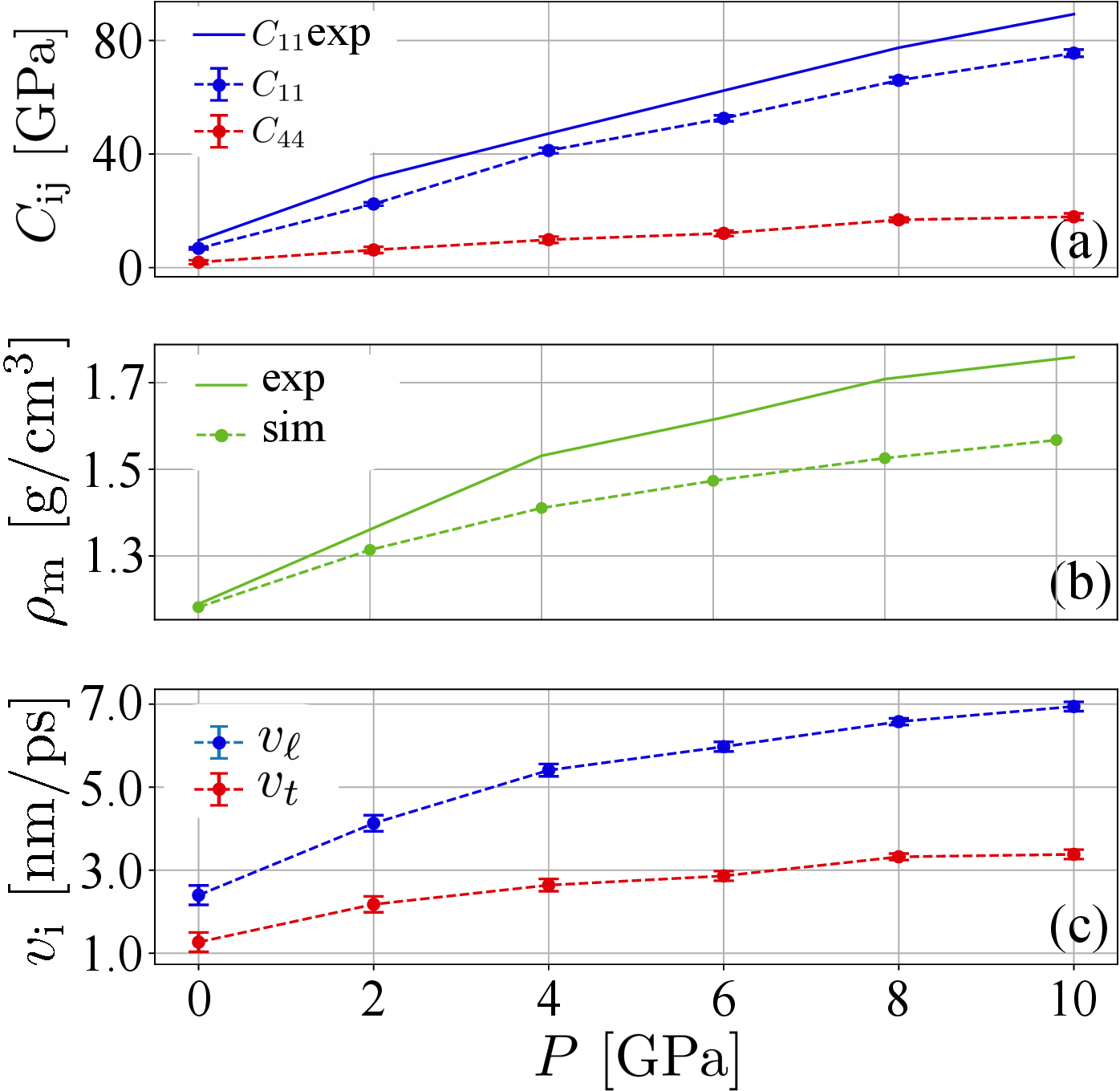}
        \caption{Pressure dependence of the components of the elastic modulus tensor $C_{\rm ij}$ (part a), mass density $\rho_{\rm m}$ (part b), and 
        components of the sound-wave velocities $v_{\rm i}$ (part c). The data is shown for a poly(methyl methacrylate) sample and at a 
        temperature of $T = 300$ K. For comparison, we have also included the available $C_{11}$ experimental data of PMMA~\cite{CahillHighP} in part a.
        Experimental data for $\rho_{\rm m}$ is also included in part b~\cite{PMMADens}.
\label{fig:stiff}}
\end{figure}

In Figure~\ref{fig:stiff}(a), we show the pressure dependence of the elastic constants $C_{11}$ and $C_{44}$, as well as the 
associated longitudinal $v_{\ell}$ and transverse $v_t$ sound velocities in Figure~\ref{fig:stiff}(c). These quantities are key inputs for understanding 
how mechanical stiffness changes with pressure, and they also directly influence thermal transport in polymeric systems~\cite{Cahill16Mac,CahillHighP}.
As pressure increases from 1 atm to 10 GPa, both $C_{11}$ and $C_{44}$ increase by nearly an order-of-magnitude. 
This strong stiffening reflects the suppression of molecular motion and enhanced intermolecular forces as the polymer 
chains are 
packed more tightly. 
In contrast, the mass density $\rho_{\rm m}$ increases by a comparatively moderate 25\% across the same pressure range. 
This indicates that while compression does densify the system, the relative change in density is much smaller than the increase in stiffness. 
It can also be appreciated that $C_{\rm ij}$ increases almost linearly, especially for the higher pressures.

\begin{figure}[ptb]
	\includegraphics[width=0.49\textwidth,angle=0]{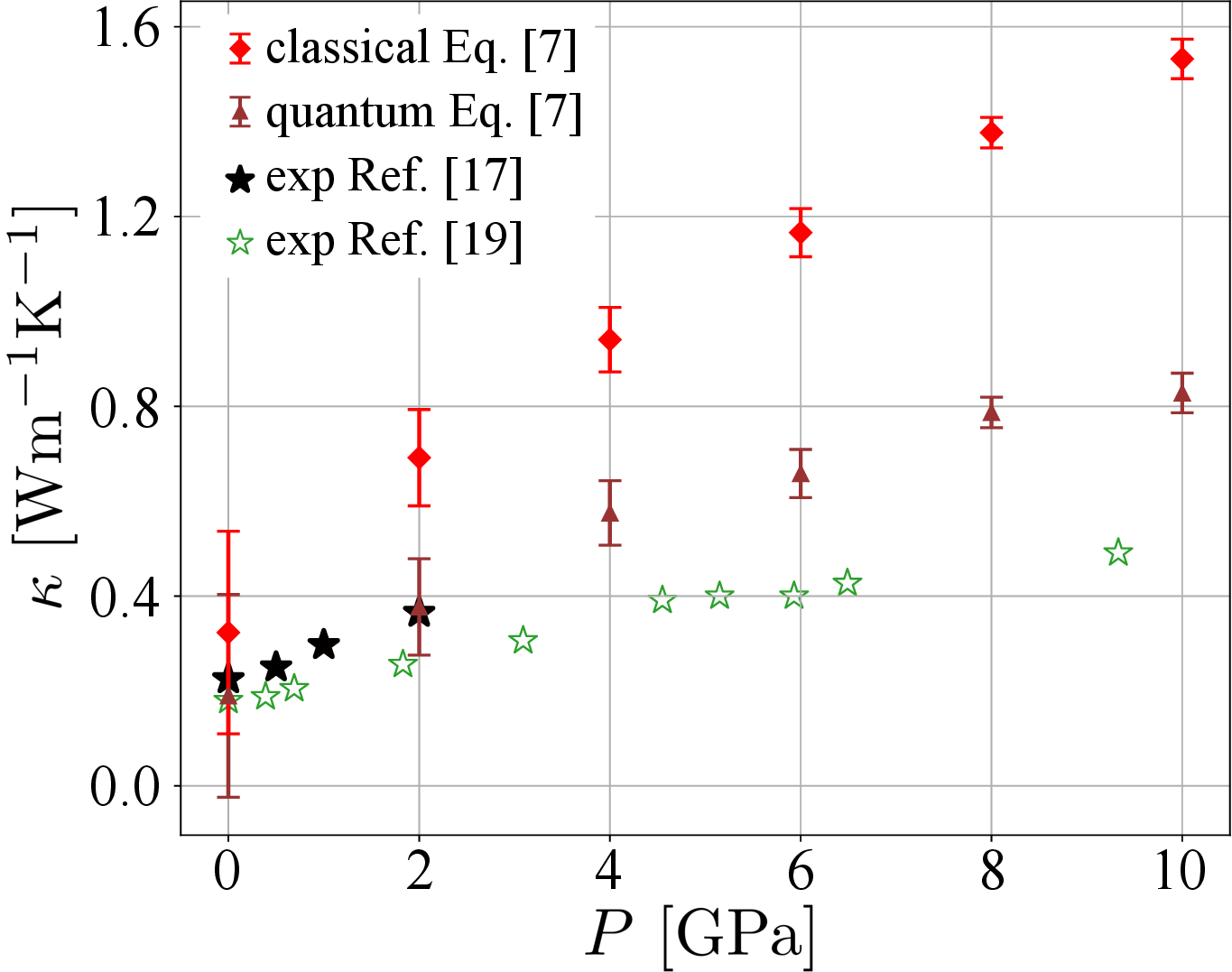}
        \caption{Thermal transport coefficient $\kappa$ as a function of pressure $P$ for poly(methyl methacrylate) (PMMA) 
        at a fixed temperature of $T = 300$ K. The data presented correspond to $\kappa$ values obtained using 
        multiple computational approaches. Classical estimates are calculated using 
        Equation~\ref{eq:mtm} in the high-temperature limit, where all vibrational modes are thermally active (solid diamonds).
        Quantum-corrected estimates of the thermal conductivity $\kappa$ are calculated using the full expression 
        given in Equation~\ref{eq:mtm} (solid triangles). 
        For comparison, experimental measurements from Ref.~\cite{Andersson1994} are included for pressures up to 2 GPa (solid stars), 
        while additional data spanning the full pressure range (open stars) are taken from Ref.~\cite{CahillHighP}.
\label{fig:kappa_quant}}
\end{figure}

Using the above computed vibrational density of states $g(\nu)$ and pressure-dependent sound velocities $v_{\rm i}$, 
we calculated the quantum-corrected thermal transport coefficient $\kappa$ at various pressures using Equation~\ref{eq:mtm}. 
These results are presented in Figure~\ref{fig:kappa_quant}. As expected, incorporating quantum corrections significantly 
improves agreement with experimental data.
In the low-pressure regime ($P \le 2.0$ GPa), the quantum-corrected $\kappa$ shows notably good agreement with the 
experimental results in Ref.~\cite{Andersson1994}, indicated by solid stars in Figure~\ref{fig:kappa_quant}. 
This consistency serves as an important validation of the modeling approach, including the methodology used for calculating $g(\nu)$
and the elastic constants, from which the sound velocities $v_{\rm i}$ are derived.
However, at higher pressure, a growing deviation appears between the simulation-based quantum-corrected $\kappa$ 
and the experimental data reported in Ref.~\cite{CahillHighP}, shown as open stars. 
This difference may arise from multiple sources:\\
(i) One potential cause lies in the pressure dependence of the mass density $\rho_{\rm m}$ in Figure~\ref{fig:stiff}(b), which plays a central role in determining 
both the sound velocities and the prefactor in Equation~\ref{eq:mtm}. As seen in Figure~\ref{fig:stiff}(b), the experimentally measured 
$\rho_{\rm m}$ may be overestimated, particularly at high pressures. Since the sound velocities are inversely proportional to 
$\sqrt {\rho_{\rm m}}$, any overestimation in density directly leads to lower values of $v_{\ell}$ and $v_t$, 
and hence smaller predicted values of $\kappa$.\\
(ii) Different experimental techniques often result in variations in reported $\kappa$ values, especially under high-pressure 
conditions where accurate control and measurement are challenging. Furthermore, the compressive effects on polymer 
microstructure -- such as chain alignment, segmental ordering, or densification 
-- may not be fully captured 
in the simulations unless large-scale rearrangements or long time scales are sampled.\\
(iii) The quantum correction in Equation~\ref{eq:mtm} assumes an isotropic material, which may become less accurate as 
pressure alters the local packing and potentially induces structural anisotropies. These assumptions might be 
violated at high pressures where subtle short-range ordering can emerge, especially in systems like polymers.\\
(iv) Although the vibrational density of states is calculated from simulations and includes anharmonic effects, its precise 
shape under the high-pressure conditions -- in particular, the high-frequency tail -- can critically influence the integral in Equation~\ref{eq:mtm}. 
Small discrepancies in this region, which is harder to sample accurately due to lower thermal population, 
may lead to noticeable differences in the final computed $\kappa$.\\
(v) Another likely source of discrepancy is the force-field itself, which is typically parameterized at ambient conditions and may not accurately capture pressure-induced changes in interatomic interactions, molecular packing, or segmental dynamics. In particular, nonbonded parameters may inadequately represent repulsive interactions at reduced intermolecular distances, and the absence of many-body or polarization effects further limits the transferability of force-field to high-pressure regimes.\\
Despite these deviations, the quantum-corrected results capture the essential physics of pressure-enhanced $\kappa$ in polymers. 
The level of agreement with at least one set of experimental data is encouraging, especially considering the 
relatively simple form of the model and the inherent complexity of amorphous polymer systems. 
Overall, this analysis supports the view that combining simulation-based structural and vibrational data with quantum-informed models 
offers a powerful and predictive framework for exploring thermal transport properties under extreme thermodynamic conditions.

\section{Conclusions}
\label{sec:conc}

In this work, we presented a comprehensive investigation into the pressure $P$ dependence of thermal transport in amorphous polymers, 
focusing specifically on poly(methyl methacrylate) (PMMA) and polylactic acid (PLA). 
Using classical molecular dynamics simulations complemented by a theoretical framework, 
we systematically quantified how pressure enhances the thermal transport coefficient $\kappa$ in these materials.

Our simulations reveal a pronounced increase in $\kappa$ with $P$, with PMMA exhibiting a nearly 3.5-fold enhancement 
and PLA a 2-fold increase as $P$ is raised from ambient conditions to 12 GPa. According to the minimum thermal conductivity model~\cite{Cahill90PRB}, this trend can be attributed to the intrinsic stiffening of the amorphous polymer under compression.
These results are consistent with and extend previous experimental findings~\cite{Andersson1994,CahillHighP}, offering both validation 
and new insight into pressure-mediated thermal behavior in amorphous polymers.

To understand the microscopic origin of this enhancement, we employed the single-chain energy-transfer model (SCETM)~\cite{MM21acsn,MM21mac}, 
which decouples energy flow along bonded and nonbonded pathways within the polymer. Our analysis demonstrates that pressure most 
significantly amplifies the nonbonded energy-transfer rate $G_{\rm nb}$ -- by nearly a factor of 6 -- as local packing increases 
and interactions become more prominent. In contrast, the transfer rate between bonded monomers $G_{\rm b}$ rises more modestly, 
by a factor of 2–3. This distinct response highlights the importance of nonbonded interactions in dictating 
thermal transport in the amorphous polymers, especially under compression. Additionally, we validated 
a simplistic expression for estimating $\kappa$ from microscopic energy-transfer rates, which shows same trends as our simulation data.

We further explored the elastic properties of these systems under pressure using a 
newly developed, noise-reducing technique for calculating the elastic constants $C_{11}$ and $C_{44}$ at finite temperatures~\cite{Muser25sub}. 
This methodology provides accurate estimates of pressure-dependent sound velocities, which are critical for the calculation of $\kappa$.

Recognizing that classical simulations tend to overestimate $\kappa$~\cite{kappaOil,Mukherji24PRM} -- 
primarily because they assume full thermal activation of all vibrational modes~\cite{MHM21prm} -- we also incorporated quantum corrections into our analysis. 
By integrating the exact vibrational density of states $g(\nu)$ into the minimum thermal conductivity framework~\cite{Mukherji24PRM}, we were able to account for the quantum 
corrections of the high-frequency vibrational modes (such as those involving stiff C-H bonds), which are inactive at room temperature. 
This approach yields a quantum-corrected estimate of $\kappa$ that are in good agreement with experimental values and 
significantly improves the physical realism of our simulations.

To summarize, our study provides a multiscale, mechanistic understanding of pressure-enhanced thermal transport in polymers. 
We bridge atomistic insights, vibrational analysis, elastic property characterization, and macroscopic transport 
modeling into a coherent picture that explains how and why $\kappa$ increases under compression. These findings not 
only deepen our fundamental understanding of heat transport in disordered polymeric materials but also offer 
practical guidance for designing polymer-based systems -- such as thermal interface materials, 
flexible electronics, and insulating composites -- where controlling thermal conductivity under variable mechanical conditions is crucial.
Overall, this study provides microscopic insights into how pressure affects thermal transport in 
polymer systems and offers guiding principles for the design of polymer-based materials suited for extreme environments.

\noindent{\bf Acknowledgment:}
J.W. and D.M. thank the ARC Sockeye facility in British Columbia Canada within the project allocation st--mukherji--1, where the majority of these simulations were performed.
We gratefully acknowledge the Gauss Centre for Supercomputing e.V. (www.gauss--centre.eu) 
for providing computing time through the John von Neumann Institute for Computing (NIC) on the 
GCS Supercomputer JUWELS at J\"ulich Supercomputing Centre (JSC), for additional compute time.
O.H.M.A. acknowledges CAPES for supporting his stay in France within a double-degree program.
Financial support for D.M. and M.M. was provided by the Bundesministerium f\"ur Technologie, Forschung und Raumfahrt (BMTRF) 
within the project 16ME0658K MExMeMo and European Union– NextGenerationEU.\\
With deep appreciation, we dedicate this work to Kurt Kremer on the occasion of his 70th birthday-- 
an inspiring colleague, cherished friend to D.M. and M.M., and a trusted mentor to D.M.

\bibliographystyle{ieeetr}
\bibliography{MukherjiarXiv}

\end{document}